\documentclass[conference]{IEEEtran}
\IEEEoverridecommandlockouts
\usepackage{cite}
\usepackage{amsmath,amssymb,amsfonts}
\usepackage{algorithmic}
\usepackage{graphicx}
\usepackage{textcomp}
\usepackage{xcolor}
\usepackage{hyperref}
\usepackage{siunitx}
\usepackage{multirow}
\usepackage{refstyle}
\def\BibTeX{{\rm B\kern-.05em{\sc i\kern-.025em b}\kern-.08em
    T\kern-.1667em\lower.7ex\hbox{E}\kern-.125emX}}
\usepackage{booktabs, caption}
\usepackage{threeparttable}

\newcommand{\refappendix}[1]{\hyperref[#1]{Appendix~\ref*{#1}}}

\DeclareCaptionLabelFormat{tablel}{#1L-#2}
\usepackage{graphicx}
\ifCLASSOPTIONcompsoc
    \usepackage[caption=false, font=normalsize, labelfont=sf, textfont=sf]{subfig}
\else
\usepackage[caption=false, font=footnotesize]{subfig}
\fi

\begin{document}

\title{Generative Adversarial Networks for photo to Hayao Miyazaki style cartoons\\}

\author{\IEEEauthorblockN{1\textsuperscript{st} Filip Andersson}
\IEEEauthorblockA{\textit{Department of computer science} \\
\textit{J\"onk\"oping University}\\
J\"onk\"oping, Sweden \\
}
\and
\IEEEauthorblockN{2\textsuperscript{nd} Simon Arvidsson}
\IEEEauthorblockA{\textit{Department of computer science} \\
\textit{J\"onk\"oping University}\\
J\"onk\"oping, Sweden \\
}
}

\maketitle


\begin{abstract}
    This paper takes on the problem of transferring the style of cartoon images to real-life photographic images by implementing previous work done by CartoonGAN.
    We trained a Generative Adversial Network(GAN) on over 60 000 images from works by Hayao Miyazaki at Studio Ghibli. To evaluate our results, we conducted a qualitative survey comparing our results with two state-of-the-art methods.
    
    117 survey results indicated that our model on average outranked state-of-the-art methods on cartoon-likeness.
    
\end{abstract}


\begin{IEEEkeywords}
    cartoon, anime, Generative Adversarial Networks, style transfer
\end{IEEEkeywords}

\section{Introduction}

Cartoons are artistic mediums, with both varying style and content. The creation of cartoons is usually time consuming and require artistic skill. The concept of style transferring neural networks in the field of machine learning can be used to automate the process of converting real life photographs into cartoon style, drawn images.

The area of style transferring neural networks can be divided into two categories: paired and unpaired. The paired approach entails providing a source image (here, real-life image) and a target style image (here, cartoon image) to be applied to the source image. This approach requires a very specific type of pre-stylized and post-stylized images. Datasets for this approach can therefore be hard to acquire. In contrast, the unpaired approach takes two separate sets of images; One set of source (real-life) images and target (cartoon) images. From the two image sets, a model is trained to produce images, keeping the source content structure but applying the target style.

A popular method for unpaired style transfer is Generative Adversarial Networks (GAN)\cite{Goodfellow_2014_GAN}. The main idea of GANs is that they allow us to generate samples from a new, unknown distribution of data, using previously seen data. It has been found that with enough data and computing power it is possible to recreate visual features with striking similarity to the target \cite{Tero_2019_styleGAN2}. Many see the possibility of using GANs to create new data when the domain's data is limited.

-Reference \cite{Chen_2018_CartoonGan} proposed a method, CartoonGAN, to automate the process of creating cartoon images. By applying a convolutional neural network to real-life photographs, and transferring the artistic style of existing cartoons to these photos. Their approach focuses on the combination of convolutional neural networks (CNN) in a combination with GANs. Their work bases itself on the previously proposed method by \cite{Gatys_2016_StyleTransfer}.

\cite{Hicsonmez_2020_GANILLA} proposed a method using GANs called GANILLA, trained on drawings from children's books but also the style of drawings by Hayao Miyazaki. They also propose a quantitative evaluation method using two CNNs to classify style and content transfer.

We aim to implement our own method based on CartoonGAN with a more extensive dataset and to compare it with CartoonGAN\cite{Chen_2018_CartoonGan} and GANILLA\cite{Hicsonmez_2020_GANILLA} using each network's pre-trained weights for Hayao Miyazaki style.
\section{Related Work}
Since their introduction in \cite{Goodfellow_2014_GAN}, Generative Adversarial Networks (GAN) have become a heavily studied and improved upon area in machine learning \cite{Zhang_2019_SAGAN, Zhu_2017_CycleGAN}.
Attempts have been made to allow the usage of GANs in image to image tasks.
\cite{Zhu_2017_CycleGAN} proposed an Encoder-Decoder architecture to create high quality style transfer using the GAN architecture. This approach allows more stable training compared to earlier approaches. \cite{Hicsonmez_2020_GANILLA} adopts this architectural design in their work.

\cite{Gatys_2016_StyleTransfer} proposed Style transfer, with the purpose of synthesising a new image using two images from separate domains, one being the content target, whereas the other contains the style wished to transfer. They propose the usage of a pretrained VGG network, where the higher level feature maps are used to extract the style of the target image, whereas the lower levels are used to extract the content. The goal being to achieve a low error rate on all levels.
\section{Implementation}

\subsection{Datasets}

For real-life images we decided to use the Flickr30K dataset \cite{Young_2014_flickr30k} which contains 31 783 images, crowd sourced from Flickr.com\footnote{\url{https://www.flickr.com/}}. The images contain annotated bounding boxes of what each image represent. For this work we discard the annotations and solely use the raw images.

For cartoon images we used works by Hayao Miyazaki to best be able to compare our method with CartoonGAN and GANILLA. The dataset contains 63 010 images from 23 different movies produced by Studio Ghibli. Using scenedetect\footnote{\url{https://pypi.org/project/scenedetect/}} we captured 3 images per movie scene (at the start, middle, and end of each scene) in a resolution of 256x256 pixels and thereafter removed images deemed too similar to each other using VisiPic\footnote{\url{http://www.visipics.info/}}.

Following CartoonGAN's implementation\cite{Chen_2018_CartoonGan}, we also created a \textit{smoothed dataset} from our cartoon dataset. This was done by first identifying edge pixels using a canny filter with a threshold between 150 and 500 and thereafter applying a Gaussian blur filter on a 3x3 pixel dilation of the identified edges. All image filter handling was done using OpenCV\cite{Bradski_2008_openCV}.

\begin{figure}[!ht]
    \begin{center}
        \noindent\includegraphics[width=0.48\textwidth, keepaspectratio=false]{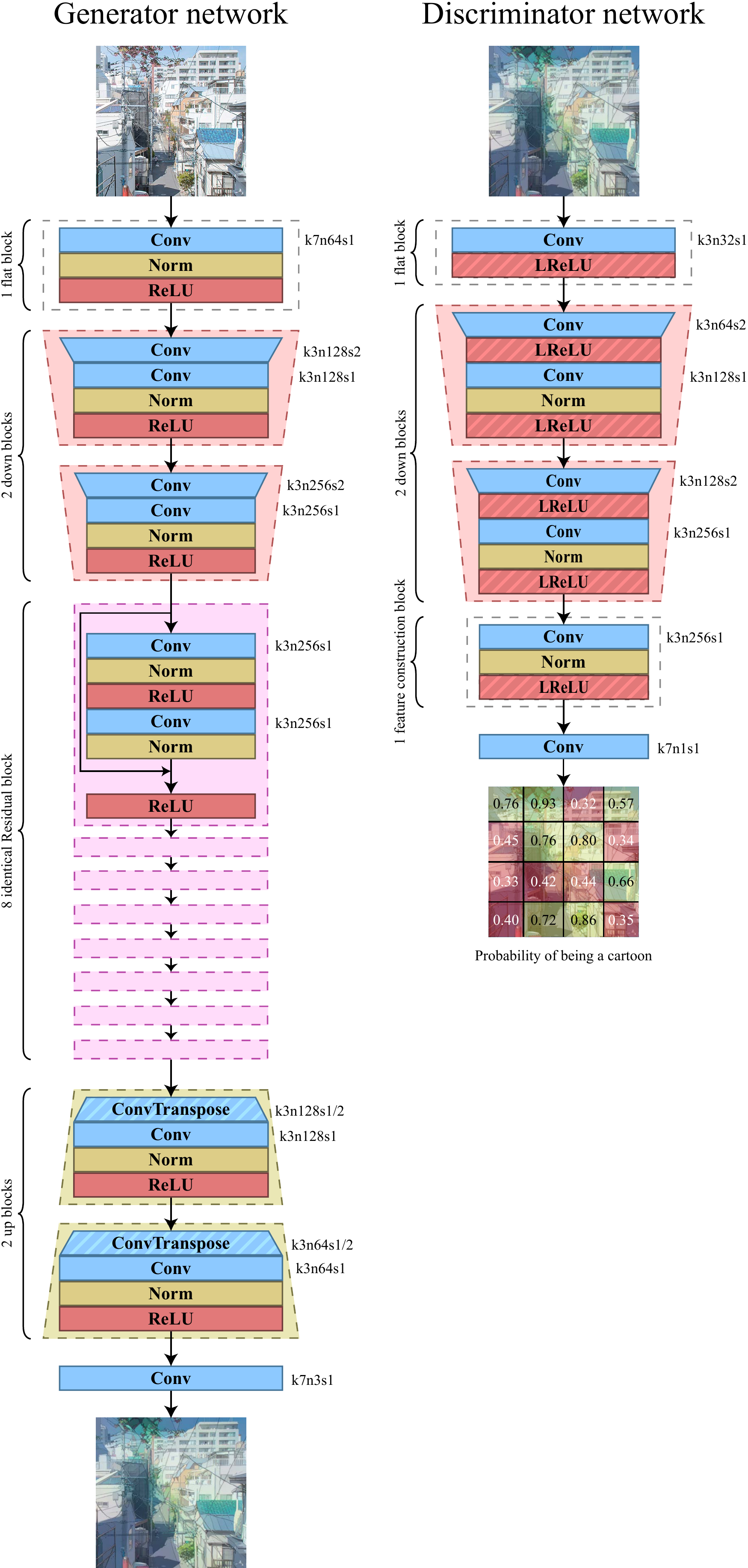}
        \caption{
        Architecture of generator(left) and discriminator(right) specifying layers: Convolutional (\textit{Conv}), Batch-Normalization (\textit{Norm}), Rectified Linear Unit (\textit{ReLU}), Leaky ReLU (\textit{LReLU}), Transposed Convolution (\textit{ConvTransposed}). Noted to the right of each convolutional layer is kernel size (\textit{k}), neurons (\textit{n}), and stride (\textit{s}).
        }
        \label{fig:Architecture}
    \end{center}
\end{figure}

\subsection{Network architecture}
\label{Network architecture}

We implemented the model using PyTorch\cite{Paszke_2019_Pytorch}. All code is available through Github (see \refappendix{app:coderepo}). Our implementation, like GANs in general, consists of a generator G and discriminator D. Both networks are implemented from the descriptions in \cite{Chen_2018_CartoonGan}. 

The discriminator is trained on both the style images and the generator results to be able to classify what images are original cartoons. The generator is trying to find a style to a apply to the source image such that its result is able to fool the discriminator. The discriminator is then trained on the results from the generator and the cycle continues. The networks are constantly competing with each other, improving with time. 

The generator network, used to transform real-life images to cartoon versions, consists of 14 layers or blocks (see \autoref{fig:Architecture}); 1 flat convolution block, 2 blocks for down-convolution, 8 residual blocks\cite{He_2016_ResNet}, 2 blocks for up-convolution, and 1 convolution layer. The residual blocks are adapted from the layout proposed in \cite{johnson_2016_Perceptual_Losses}.

The discriminator network, used to classify an image as either cartoon or real-life, consists of 5 layers; 1 flat convolution layer, 2 down-convolutions, 1 feature construction block, and 1 convolution layer. The discriminator is what is called a \textit{patchGAN}\cite{Li_2016_PatchGan} which means that rather than giving a single classification per image, it is instead used on smaller, cropped sections (patches) of each image to produce a list of classifications results, one for each patch. Here we use Leaky ReLU (LReLU) \cite{Maas_2013_Rectifier_Nonlinearities} with $\alpha$ = 0.2 after every normalization as per \cite{Chen_2018_CartoonGan}.

\subsection{Loss functions}

A loss function is used to evaluate the error, and therefore the performance of a model. In the context of GANs it is used to evaluate the performance of both networks \cite{Goodfellow_2014_GAN}. \cite{Chen_2018_CartoonGan} proposes the use of a combined loss function $L(G, D)$ which takes into account both adversarial loss $L_{adv}(G, D)$ and content loss $L_{con}(G, D)$ for both networks G and D (see \eqref[name=Eq.~]{eqn:loss_function}).

\begin{equation}\eqlabel{eqn:loss_function}
    L(G, D) = L_{adv}(G, D) + \omega L_{con}(G, D)
\end{equation}

$\omega$ is a scalar used to balance the two loss functions. Per \cite{Chen_2018_CartoonGan}, we set $\omega$ = 10.

The adversarial loss can be seen as a measure of how well the combined networks can transfer the cartoon style to the target image whereas the content loss can be seen as a measure of how well the content structure of the source image is preserved. Optimizing the adversarial loss is a \textit{minimax problem} \cite{Kjeldsen_2001_Minimax} where the discriminator network tries minimize the chance to do a wrong prediction, whereas the generator network aims to maximize the probability of fakes that the discriminator misclassifies as real.

\section{Experiment}
\subsection{Model training}
\label{modelTraining}
The training was conducted using a NVIDIA RTX 2080 Graphics card. Each image was resized to a size of 224x224 pixels. We replicated the initialization phase done by \cite{Chen_2018_CartoonGan} and trained it for 10 epochs. Each iteration was trained on a batch of 11 images. The optimizer used was AdamW \cite{Loshchilov_2019_ADAMW} which is an improvement on \cite{Kingsma_2015_Adam}. The learning rate was set to a constant \num{1e-3} and a weight decay of \num{1e-4}. We also used a cyclic learning rate scheduler \cite{L_2017_CyclicLR} with the maximum learning rate set to \num{1e-2}. Both networks were trained for a total of 60 epochs using the entire flickr30k dataset\cite{Young_2014_flickr30k}.

\subsection{Evaluation}
Evaluation of GANs can be a complicated area, especially in the context of style transfer where the result is an altered image. \cite{Hicsonmez_2020_GANILLA} proposed a quantitative evaluation where two neural networks are used to evaluate both how well the content of the source image is preserved (Content-CNN) and how well the style of the target image is transferred (Style-CNN) respectively. They also propose a method for qualitative evaluation, comparing 4 different style transfer GANs to identify which model has the best "style and content identifiability"\cite{Hicsonmez_2020_GANILLA}.

The accuracy and merit of using CNNs to evaluate results in a quantitative manner is questionable. The discriminator described in \autoref{fig:Architecture} is very similar to the Style-CNN of \cite{Hicsonmez_2020_GANILLA}, both applying patchGAN\cite{Li_2016_PatchGan}. Taken into account that the discriminator already evaluates on both style and content, we have chosen to only perform a qualitative evaluation using human subjects.

We aim to compare 3 different implemented models\footnote{CartoonGAN implementation: \url{https://github.com/Yijunmaverick/CartoonGAN-Test-Pytorch-Torch}}$^{,}$\footnote{GANILLA implementation: \url{https://github.com/giddyyupp/ganilla}} of cartoon style transfer GANs: CartoonGAN\cite{Chen_2018_CartoonGan}, GANILLA\cite{Hicsonmez_2020_GANILLA} and our method. For all models we use weights trained on artwork by Hayao Miyazaki. To compare the 3 models we designed a survey. For each task, we showed the participant 3 images, one produced by each different model from a common source image. The participant then had to rank the 3 images on a scale from 1-3 based on a certain statement. Here 1 agrees the most with the given statement and 3 agrees the least. For the questions, we asked the participant to "rank the following images according to aesthetic/visual appeal"(henceforth referred to as \textit{Aestheticness}) and to "rank the following images according to how much the image resembles a cartoon (i.e. illustrated photo or anime)"(henceforth referred to as \textit{Cartooniness}).

\begin{figure}[ht]
    \centering
    \subfloat[Survey part 1\figlabel{fig:survey_a}]{%
       \includegraphics[width=0.49\linewidth]{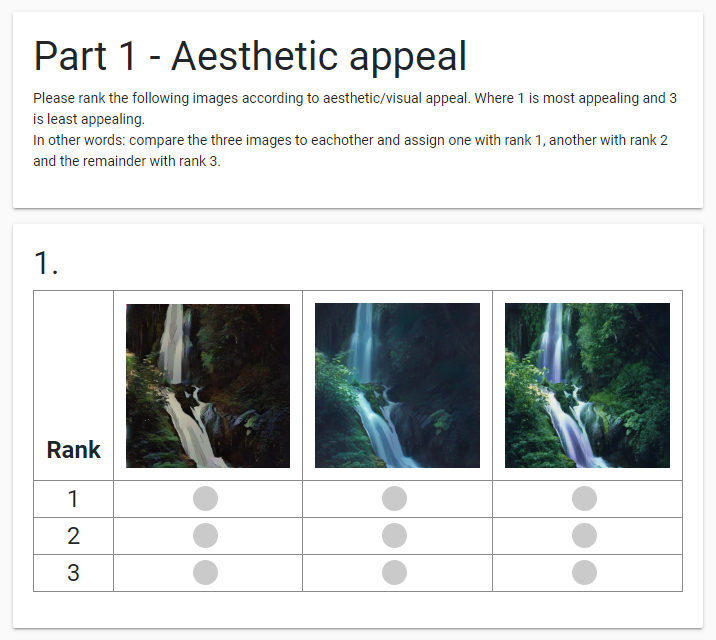}}
    \hfill
    \subfloat[Survey part 2\figlabel{fig:survey_b}]{%
        \includegraphics[width=0.49\linewidth]{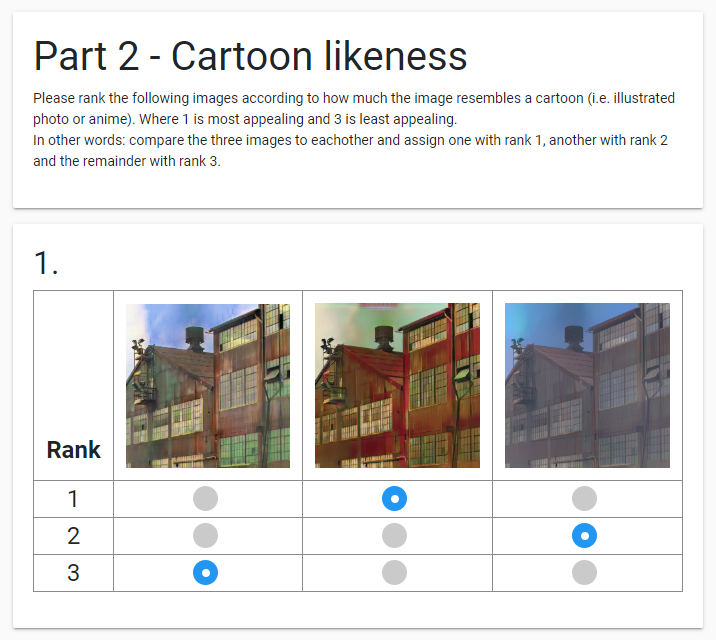}}
    \caption{The survey questions and examples of tasks. (a) First question part and its first random task. (b) Second question part and its first random task with marked answers.}
    \label{fig:Survey}
\end{figure}

To survey participants, we built a simple web page hosted on GitHub Pages\footnote{\url{https://pages.github.com/}} and collected results using Pageclip\footnote{\url{https://pageclip.co/}}. We used a total of 20 source images for the survey, 10 for each question, one for each task. The order of the images (and hence tasks) were shuffled across both questions for each participant. In this way every source image receives answers on both questions. Examples of the survey views can be seen in \autoref{fig:Survey}.

The 20 source images were obtained from from Unsplash.com\footnote{\url{https://unsplash.com/}}. These images as well as the results from CartoonGAN, GANILLA, and our method can be found on the code repository page (see \refappendix{app:coderepo}).
\section{Results}

\begin{figure*}[!ht]
    \begin{center}
        \noindent\includegraphics[width=\textwidth]{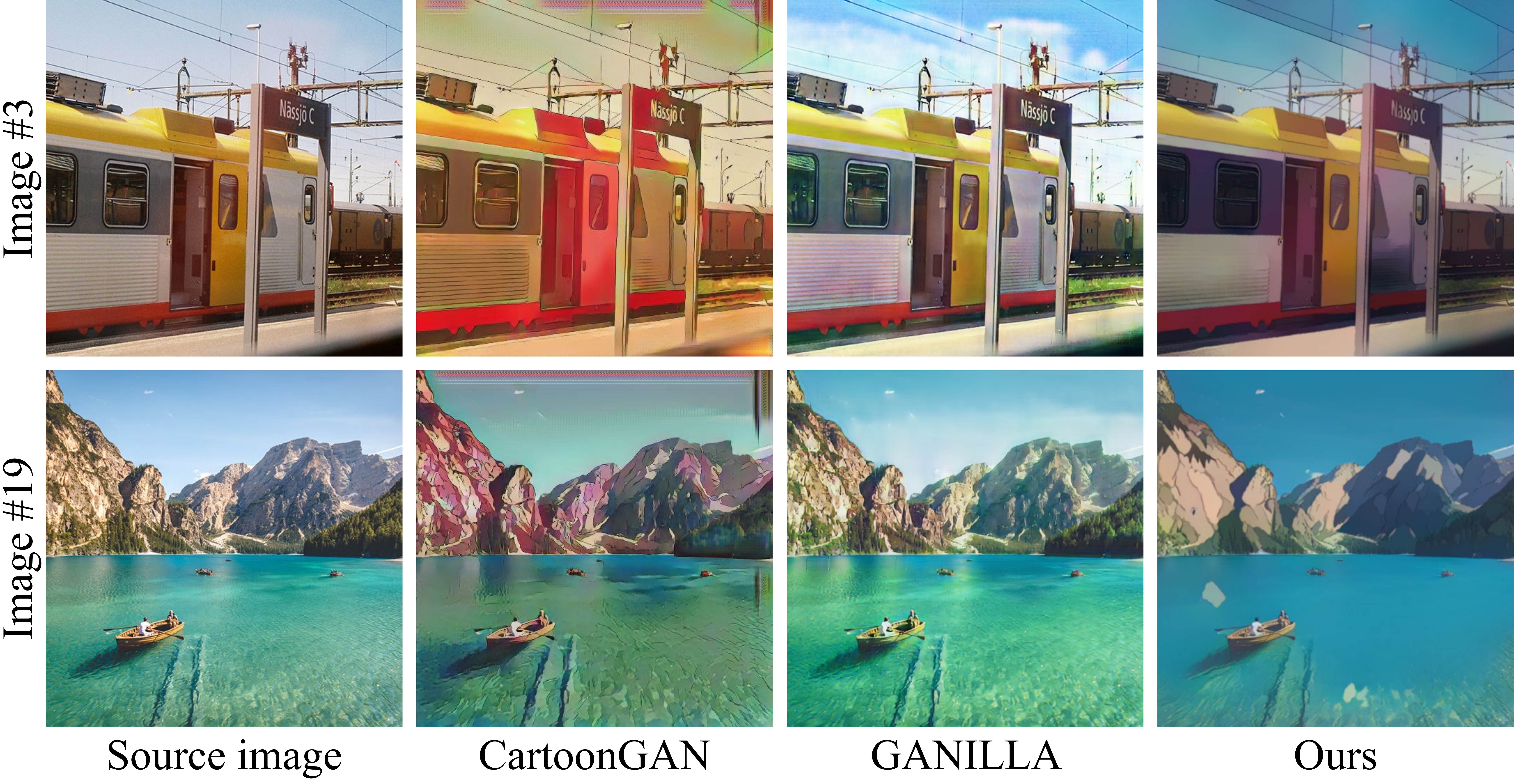}
        \caption{Results of two sample source images stylized by CartoonGAN\cite{Chen_2018_CartoonGan}, GANILLA\cite{Hicsonmez_2020_GANILLA} and our method respectively.}
        \label{fig:comparisonGrid}
    \end{center}
\end{figure*}

\begin{table}[htbp]
    \caption{Mean rank of survey results for each model}
    \label{tab:surveyResults}
    \begin{center}
        \begin{tabular}{|c|c|c|c|}
            \hline
            \multirow{2}{*}{\textbf{Question}} & \multicolumn{3}{c|}{\textbf{Model}} \\
            \cline{2-4} 
             & \textbf{\textit{CartoonGAN}} & \textbf{\textit{GANILLA}}& \textbf{\textit{Ours}} \\
            \hline
            \textbf{\textit{Aestheticness}}    & 2.12 & \textbf{1.64} & 2.24 \\ \hline
            \textbf{\textit{Cartooniness}}     & 1.90 & 2.33 & \textbf{1.78} \\ \hline
        \end{tabular}
    \end{center}
\end{table}

Two sample results of our model applied to to images can be seen in \autoref{fig:comparisonGrid} (leftmost column). These images, together with the images produced by CartoonGAN and GANILLA also shown in \autoref{fig:comparisonGrid} (middle columns) made up two of the 20 tasks of the survey.

The survey had a total of 117 participants, each providing 60 survey inputs. The mean score of can be seen in \autoref{tab:surveyResults} where bold marks the highest score for each question.
\section{Analysis}


From the sample results shown in \autoref{fig:comparisonGrid}, we can see that both CartoonGAN and our model produces images with more flat surfaces and a mono color tint applied. CartoonGAN's results seem to have a red tint whereas our results seem to have a dark blue tint. The difference is most likely due to the difference in datasets used. GANILLA on the other hand often produces more saturated images that more resembles the source image. The lack of cartoonization here is most likely due to the limited dataset used; As stated in GANILLA's implementation for the Hayao Miyazaki model: 

\begin{quote}
    [...] we were not able to replicate the results of CartoonGAN on our collection due to the low quality of the samples. In order to be able to perform experiments on the same illustrator with CartoonGAN, instead, we have collected images of Miyazaki Hayao using Google Image Search to be used as our target training set.
\end{quote}

They go on to say:

\begin{quote}
    [...] this is a more challenging dataset compared to the Hayao dataset used in \cite{Chen_2018_CartoonGan} which was composed of a unique style from a single story (Spirited Away). Our images correspond to the samples from the entire collection of Hayao and therefore mixture of several styles from a variety of stories.
\end{quote}

This also applies to our model which most likely uses a larger dataset than both \cite{Chen_2018_CartoonGan} and \cite{Hicsonmez_2020_GANILLA} but is still able to reproduce the results of CartoonGAN.

As seen in \autoref{tab:surveyResults}, our model received on average a worse score on aestheticness compared to the other two models. GANILLA received the best aestheticness score with CartoonGAN and our implementation performing at similar levels. This is reasonable since our implementation is based on CartoonGAN and share very similar implementation and network architecture. One potential reason for GANILLA's high rank on aestheticness could be the fact that by not cartoonizing images to a high degree, this model also avoids the creation of any potential artifacts that can make the result less appealing. People might also find cartoonized images less appealing in general.

When it comes to cartooniness, our model received the best score out of the three models, with CartoonGAN ranking close yet again. As mentioned above, GANILLA was not able to produce very cartoonized images, performing best at images of nature and landscapes.

\begin{figure*}[htbp!]
    \minipage[t]{0.19\textwidth}
        \subfloat[][\centering Eye anomalies (from image \#18)\label{fig:anomaly_a}]{
        \includegraphics[width=0.96\textwidth]{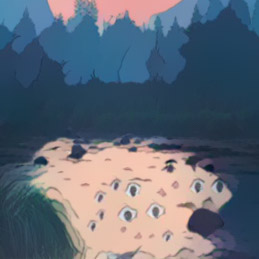}}
    \endminipage\hfill
    \minipage[t]{0.19\textwidth}
        \subfloat[][\centering Hair and eye anomalies (from image \#19)\label{fig:anomaly_b}]{
        \includegraphics[width=0.96\textwidth]{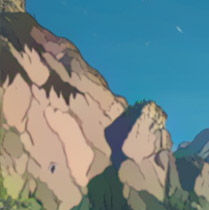}}
    \endminipage\hfill
    \minipage[t]{0.19\textwidth}
        \subfloat[][\centering Blob anomalies (from image \#8, with increased brightness)\label{fig:anomaly_c}]{
        \includegraphics[width=0.96\textwidth]{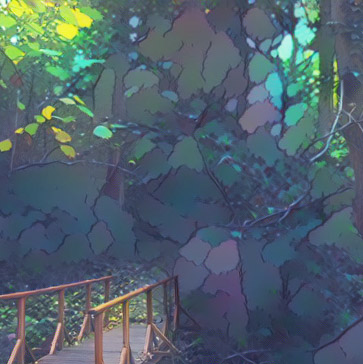}}
    \endminipage\hfill
    \minipage[t]{0.19\textwidth}
        \subfloat[][\centering Solid colors in between lines (from image \#3)\label{fig:anomaly_d}]{
        \includegraphics[width=0.96\textwidth]{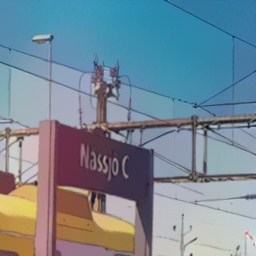}}
    \endminipage\hfill
    \minipage[t]{0.19\textwidth}
        \subfloat[][\centering Correct eye placement (from image \#10)\label{fig:anomaly_e}]{
        \includegraphics[width=0.96\textwidth]{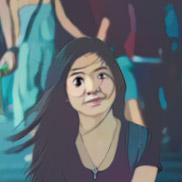}}
    \endminipage
    \caption{Interesting elements produced by our model.}
    \label{fig:anomalies}
\end{figure*}

A double edge sword of our model is the creation of homogeneously colored regions. This is a feature of cartoons in general and can look good when applied correctly (i.e. in \autoref{fig:anomaly_b}) but the model sometimes inserts these regions inappropriately, making them appear as mosaic like structures or blobs that stand out from the rest of the image (see \autoref{fig:anomaly_c}). Similarly, the model tries to smooth out color segments separated by borders. This can be seen in \autoref{fig:anomaly_d} where parts of the sky are colored in different shades when separated by overhead train wires. The same effect can be observed in \autoref{fig:anomaly_e} where the woman's skin has different colors in segments separated by hair or necklace. Homogeneous color regions like the mentioned examples does not seem to appear to as large extent in \cite{Chen_2018_CartoonGan} or \cite{Hicsonmez_2020_GANILLA}.

Further interesting interesting elements produced by our model are cartoon facial features such as eyes, mouths, and hair. \autoref{fig:anomaly_e} shows a correct placement of an eye onto a photo of a woman, only replacing the left eye while missing the right. Moreover, the model also places cartoon facial features on other skin colored surfaces that are not human faces. Examples of this can be seen in \autoref{fig:anomaly_a} where a slightly skin colored reflection of the sky had eyes and mouths applied. Similarly, in \autoref{fig:anomaly_b}, the hills have eyes whereas the shrubs or moss on top of the hills are most likely being recreated from cartoon hair. This could be a side effect of the large amount of foreground content in the cartoon dataset. \cite{Chen_2018_CartoonGan} handpicked images as to attain a more balanced dataset of both foreground and background images.


\section{Future work}
More work could be done refining the dataset. Removing images containing large amount of foreground content could result in lower amounts of facial features being applied in general. It can also be interesting to exclude foreground content completely to only transfer the style of the backgrounds i.e. landscapes. The edge smoothing of the cartoon dataset used in preprocessing could also be optimized to more general thresholds or even individual thresholds per image. These proposed refinements could in turn also be achieved by training a model, especially when dealing with as large a dataset as in our case, where manual refinement is not feasible.

Additional experiments could be conducted using different artists or studios to evaluate the reproducibility of \cite{Chen_2018_CartoonGan}. An interesting artist is Makoto Shinkai at the studio CoMix Wave Films which have produced a large discography of popular movies and is used as training data in \cite{Chen_2018_CartoonGan}.

Further, it would be interesting to examine other architectures together with the \cite{Chen_2018_CartoonGan} methodology. An interesting network architecture would be U-Net\cite{Ronneberger_2015_UNET} allowing the network to use long shortcut connections. \cite{Hicsonmez_2020_GANILLA} uses the U-Net architecture for their model. Combining the U-Net architecture with the methodology used by us and CartoonGAN, together with a large dataset could produce more consistent results.

Lastly, more time could be used to train the network. This experiment was limited regarding computational resources and time. We had no more resources than the single graphics card mentioned in \autoref{modelTraining}.
\section{Conclusion}
We attempted to reproduce the methodology of CartoonGAN\cite{Chen_2018_CartoonGan} with minor tweaks and a larger dataset collected from movies made by Hayao Miyazaki. We conducted a qualitative survey comparing our implementation to CartoonGAN\cite{Chen_2018_CartoonGan} and GANILLA\cite{Hicsonmez_2020_GANILLA}, all trained on Hayao Miyazaki images. The survey results showed that our implementation outperformed the other models in cartoon-likeness although it performed worse than the other models on aesthetical appeal.

Our model's results showed interesting elements not visible in the compared models' results. The elements include homogeneously colored areas and characteristical cartoon facial features.

Going further, interesting improvements entails refining the cartoon dataset and experimenting with more persistent architectures.

\section*{Project activities}

\begin{table}[h!]
    \caption{Time estimation for activities performed.}
    \label{tab:activities}
    \begin{center}
    \begin{threeparttable}
        \begin{tabular}{|l|c|}
            \hline
            \textbf{Activity} & \textbf{Time spent (in hours)} \\
            \hline 
            Dataset preparation & 60 \\ \hline
            Creation of survey & 70 \\ \hline
            Results analysis & 12 \\ \hline
            Model implementation & 96 \\ \hline
            Model training\tnote{a} & 144 \\ \hline
            Report writing & 80 \\ \hline
        \end{tabular}
        \begin{tablenotes}\scriptsize
            \item[a] Estimated time training the model (GPU time). Other activities were worked on in parallel.
        \end{tablenotes}
    \end{threeparttable}
    \end{center}
\end{table}

\bibliographystyle{IEEEtran}
\bibliography{references}

\appendices

\section{Code Repository}
\label{app:coderepo}
\href{https://github.com/FilipAndersson245/cartoon-gan.git}{https://github.com/FilipAndersson245/cartoon-gan.git}

\end{document}